
\input harvmac

\def\href#1#2{{#2}}

\Title{hep-th/9507135 YCTP-P12-95}{\vbox{\centerline{Extended superspace,
higher derivatives} \vskip2pt \centerline{and $SL(2, Z)$ duality}}}

\centerline{M{\aa}ns Henningson}
\bigskip{\it \centerline{Yale University}\centerline{Department of
Physics}\centerline{P. O. Box 208120}\centerline{New Haven, CT 06520-8120,
USA}}\centerline{mans@genesis5.physics.yale.edu}

\vskip 40mm \centerline{\bf Abstract}We consider the low-energy effective
action for the Coulomb phase of an $N = 2$ supersymmetric gauge theory with a
rank one gauge group. The $N = 2$ superspace formalism is naturally invariant
under an $SL(2, {\bf Z})$ group of duality transformations, regardless of the
form of the action. The leading and next to leading terms in the long distance
expansion of the action are given by the holomorphic prepotential and a real
analytic function respectively. The latter is shown to be modular invariant
with respect to $SL(2, {\bf Z})$.

\Date{July 1995, revised October 1995}

\newsec{Introduction}

In the last year and a half, there has been dramatic progress in our
understanding of strongly coupled supersymmetric gauge theories in four
dimensions. (See for example the recent review by Seiberg \ref\Seiberg{
N. Seiberg, preprint hep-th/9506077, to appear in the proceedings of PASCOS 95
and the Oskar Klein lectures.
}.)
In particular, Seiberg and Witten \ref\SW{
N. Seiberg and E. Witten, {\it Nucl. Phys.} {\bf B426} (1994) 19; {\it Nucl.
Phys.} {\bf B431} (1994) 484.
} were able to determine the metric on the Coulomb branch of the moduli space
of vacua of $N = 2$ supersymmetric $SU(2)$ gauge theories with various matter
content.

The light degrees of freedom on the Coulomb branch of such theories constitute
an $N = 2$ vector multiplet, i.e. an $N = 1$ vector multiplet and an $N = 1$
uncharged chiral multiplet. The method used in \SW\ and many subsequent papers
was to study the terms in the low-energy effective action for these fields with
at most two space-time derivatives or four fermions. These terms are determined
by a holomorphic prepotential. An $SL(2, {\bf Z})$ group of duality
transformations, acting linearly on the $N = 1$ chiral superfield and the first
derivative of the prepotential and by electric-magnetic duality on the $N = 1$
vector field, plays an important role in determining the properties of the
model.

However, the prepotential terms are just the leading terms in a systematic
long-distance expansion of the low-energy effective action. The object of this
paper is to study the exact expansion, in particular the next to leading terms.
For simplicity, we will only consider the case of a rank one gauge group; the
generalization to other groups should be straightforward.

It is convenient to work in a manifestly $N = 2$ supersymmetric formalism, and
in section two we give a quick review of $N = 2$ superspace and in particular
the $N = 2$ vector multiplet. In section three, we show that  this formalism is
naturally invariant under a group of duality transformations isomorphic to
$SL(2, {\bf Z})$, regardless of the form of the action. In section four, we
discuss the long-distance expansion of the low-energy effective action in $N =
1$, $N = 2$ and $N = 4$ supersymmetric theories. In the case of $N = 2$
supersymmetry, the leading and next to leading terms are given by the
holomorphic prepotential and a real analytic function respectively. In section
five, we show how these objects transform under $SL(2, {\bf Z})$. For the
prepotential, we recover the results of \SW\ that its first derivative and the
fundamental field transform in the defining representation of $SL(2, {\bf Z})$.
The real analytic function is found to be  modular invariant with respect to
$SL(2, {\bf Z})$.

\newsec{The $N = 2$ superspace formalism}

We will be using the formalism of Grimm, Sohnius and Wess \ref\GSW{
R. Grimm, M. Sohnius and J. Wess, {\it Nucl. Phys.} {\bf B133} (1978) 275.
}.
The $N = 2$ superspace has bosonic coordinates $x^\mu$ and fermionic
coordinates $\theta^\alpha{}_i$ and $\bar{\theta}^{\dot{\alpha} i}$, where $\mu
= 0, \ldots, 3$ is a space-time vector index, $\alpha = 1, 2$ and $\dot{\alpha}
= \dot{1}, \dot{2}$ are Weyl and anti-Weyl spinor indices respectively, and $i
= 1, 2$ indexes a doublet under the $SU(2)_R$ algebra which is part of the $N =
2$ supersymmetry algebra.

Infinitesimal supersymmetry transformations are generated by $Q_\alpha{}^i$ and
$\bar{Q}_{\dot{\alpha} i}$ defined by
\eqn\susygenerators{
\eqalign{
Q_\alpha{}^i & = {\partial \over \partial \theta^\alpha{}_i} - i (\sigma^\mu
\bar{\theta}^i)_\alpha \partial_\mu \cr
\bar{Q}_{\dot{\alpha} i} & = - {\partial \over \partial
\bar{\theta}^{\dot{\alpha} i}} + i (\theta_i \sigma^\mu)_{\dot{\alpha}}
\partial_\mu. \cr
}
}
They fulfill the algebra $\{ Q_\alpha{}^i , Q_\beta{}^j \} = \{
\bar{Q}_{\dot{\alpha i}} , \bar{Q}_{\dot{\beta} j} \} = 0$, $\{Q_\alpha{}^i ,
\bar{Q}_{\dot{\beta} j} \} = 2 \sigma^\mu{}_{\alpha \dot{\beta}} \delta^i_j
P_\mu$, where $P_\mu$ is the space-time translation operator. We will also have
use for the super-covariant derivatives $D_\alpha{}^i$ and
$\bar{D}_{\dot{\alpha} i}$ defined by
\eqn\susyderivatives{
\eqalign{
D_\alpha{}^i & = {\partial \over \partial \theta^\alpha{}_i} + i (\sigma^\mu
\bar{\theta}^i)_\alpha \partial_\mu \cr
\bar{D}_{\dot{\alpha} i} & = - {\partial \over \partial
\bar{\theta}^{\dot{\alpha} i}} - i (\theta_i \sigma^\mu)_{\dot{\alpha}}
\partial_\mu. \cr
}
}
They anti-commute with the supersymmetry generators \susygenerators\ and
fulfill
the algebra
\eqn\Dalgebra{
\eqalign{
\{ D_\alpha{}^i , D_\beta{}^j \} & = \{ \bar{D}_{\dot{\alpha i}} ,
\bar{D}_{\dot{\beta} j} \} = 0 \cr \{D_\alpha{}^i , \bar{D}_{\dot{\beta} j} \}
& = - 2 i \sigma^\mu{}_{\alpha \dot{\beta}} \delta^i_j \partial_\mu . \cr
}
}

Any superfield integrated over all of superspace with the measure $d^4 \theta
d^4 \bar{\theta}$ transforms into a total space-time derivative under the
supersymmery transformations generated by \susygenerators. The same applies to
a chiral (anti-chiral) superfield, i.e. a superfield annihilated by
$\bar{D}_{\dot{\alpha} i}$ \ ($D_\alpha{}^i$), integrated over chiral
(anti-chiral) superspace with the measure $d^4 \theta$ \ ($d^4 \bar{\theta}$).

The $N = 2$ vector multiplet may be described by a complex superfield $A$ in
the adjoint representation of the gauge group which fulfills a chirality
constraint
\eqn\chiral{
\bar{D}_{\dot{\alpha} i} A  = 0
}
and a Bianchi identity constraint
\eqn\Bianchi{
D^{\alpha i} D_\alpha{}^j A = \bar{D}_{\dot{\alpha}}{}^i \bar{D}^{\dot{\alpha}
j} \bar{A}.
}
In terms of $N = 1$ superfields, $A$ contains a vector field strength
$W_\alpha$ and an $N = 1$ chiral field $\Phi$ obeying the usual constraints
$\bar{D}_{\dot{\alpha}} W_\alpha = 0$, $D^\alpha W_\alpha =
\bar{D}_{\dot{\alpha}} \bar{W}^{\dot{\alpha}}$ and $\bar{D}_{\dot{\alpha}} \Phi
= 0$. (Our conventions for $N = 1$ superspace are those of Wess and Bagger
\ref\WB{
J. Wess and J. Bagger, {\it Supersymmetry and Supergravity}, (Princeton
University Press 1983).
}.)
Decomposing further, we see that the component fields are a gauge potential
$A_\mu$, a complex scalar $\phi$, an $SU(2)_R$ doublet $\lambda_\alpha{}^i$ of
Weyl fermions and an $SU(2)_R$ triplet $E^A$, $A = 1, 2, 3$, of auxiliary
fields.

In the abelian case, the constraints \chiral\ and \Bianchi\ were solved by
Mezincescu \ref\Mezincescu{
L. Mezincescu, JINR report P2-12572 (1979).
}
in terms of a real and symmetric but otherwise unconstrained superfield $V_{i
j}$ as
\eqn\mezincescu{
A = \bar{D}^4 D^{\alpha i} D_\alpha{}^j V_{i j} .
}

\newsec{The duality transformations}

An important fact about $N = 2$ supersymmetric gauge theories is that they in
general have a moduli space of inequivalent vacuum states. We will be
considering the Coulomb branch of this moduli space, where the gauge group is
spontaneously broken down to its maximal abelian subgroup. For simplicity of
notation, we will only consider the case of a rank one gauge group; the
generalization to larger groups should be straightforward.

The light degrees of freedom at a generic point on the Coulomb branch
constitute an $N = 2$ $U(1)$ vector multiplet $A$, and the physics is
described at low energies by some effective action functional ${\cal S}[A,
\bar{A}]$. (This low energy effective action is in general not renormalizable,
although the underlying microscopic theory must be so.)
The partition function of the model is then given by
\eqn\partition{
{\cal Z} = \int {\cal D} V_{i j} \exp i {\cal S}[A, \bar{A}].
}
We emphasize that the quantity $A$ in this expression is not an independent
field, but given in terms of the real but otherwise unconstrained field
$V_{ij}$ as in \mezincescu.

In this section, we will show that this formalism is naturally invariant under
a group of duality transformations isomorphic to $SL(2, {\bf Z})$. These
transformations are not symmetry transformations in the traditional sense, but
rather relate different descriptions (i.e. different action functionals of
different fields) of the same theory. The group $SL(2, {\bf Z})$ can be thought
of as generated by two elements $S$ and $T$ with the relations
\eqn\relations{
\eqalign{
S^2 & = 1 \cr
(S T)^3 & = 1, \cr
}
}
where $1$ denotes the identity element.

We will first describe the action of the $S$-transformation on the model. To
this end, we begin by rewriting the partition function \partition\ by letting
$A$ be a
chiral but otherwise unconstrained field (i.e. not necessarily obeying the
Bianchi constraint \Bianchi). This constraint is then enforced by means of a
real and symmetric Lagrange multiplier field
$\tilde{V}_{i j}$. We thus write the partition function as
\eqn\multiplier{
{\cal Z} = \int {\cal D} A \, {\cal D} \bar{A} \, {\cal D} \tilde{V}_{i j} \exp
i \left( S[A, \bar{A}] + \int d^4 x d^4 \theta d^4 \bar{\theta} \tilde{V}_{i j}
(D^{\alpha i} D_\alpha{}^j A - \bar{D}_{\dot{\alpha}}{}^i \bar{D}^{\dot{\alpha}
j} \bar{A}) \right).
}
Performing the path integral over $\tilde{V}_{i j}$ gives rise to a delta
functional which enforces the constraint \Bianchi. The general solution to this
constraint is paramerized by a real but otherwise unconstrained field $V_{ij}$
as in \mezincescu, so the partition functions \partition\ and \multiplier\ are
indeed equal. (The path-integral measures in these expressions are determined,
up to an overall factor, by $N = 2$ supersymmetry.) The $S$-transformation now
amounts to instead integrating out $A$ (and its complex conjugate $\bar{A}$)
from \multiplier. The partition function is then given in the dual form
\eqn\dual{
{\cal Z} = \int {\cal D} \tilde{V}_{i j} \exp i \tilde{\cal S}[\tilde{A},
\bar{\tilde{A}}],
}
where $\tilde{A} = \bar{D}^4 D^{\alpha i} D_\alpha{}^j \tilde{V}_{i j}$ and the
dual action $\tilde{\cal S}[\tilde{A}, \bar{\tilde{A}}]$ is given by
\eqn\Saction{
\exp i \tilde{\cal S}[\tilde{A}, \bar{\tilde{A}}] = \int {\cal D} A \, {\cal D}
\bar{A} \exp i \left( {\cal S}[A, \bar{A}] + \int d^4 x d^4 \theta A \tilde{A}
+ {\rm c. c.} \right).
}
Here we have used that the factor $d^4 \bar{\theta}$ in the superspace
integration measure can be replaced by $\bar{D}^4$ and analogously for the
complex conjugate. We see that the dual expression \dual\ for the partition
function is of the same form as the original expression \partition, but with a
new action $\tilde{\cal S}$ which is a functional of the new variable
$\tilde{V}_{ij}$ (through the quantities $\tilde{A}$ and $\bar{\tilde{A}}$).
The relationship \Saction\ states that the $S$-transformation acts as a Fourier
transformation in field space on the exponentiated action of the model.

We now turn to the $T$-transformation, which simply consists of adding the
terms
\eqn\thetaterm{
\int d^4 \theta {1 \over 2} A A + \int d^4 \bar{\theta} {1 \over 2} \bar{A}
\bar{A} = {1 \over 32 \pi} \epsilon^{\mu \nu \rho \sigma} F_{\mu \nu} F_{\rho
\sigma}
}
to the Lagrangian of the model. This amounts to shifting the $\theta$-angle by
$2 \pi$ and thus gives an equivalent description of the theory. The new action
is still a function of the same independent field $V_{ij}$ (through the
quantities $A$ and $\bar{A}$).

To establish $SL(2, {\bf Z})$ invariance, we must check the relations
\relations. Two consecutive $S$-transformations transform an action functional
${\cal S}$ first into $\tilde{\cal S}$ given by \Saction, and then into
$\tilde{\tilde{\cal S}}$ given by
\eqn\SS{
\eqalign{
\exp i \tilde{\tilde{\cal S}}[\tilde{\tilde{A}}, \bar{\tilde{\tilde{A}}}] & =
\int {\cal D} \tilde{A} \, {\cal D} \bar{\tilde{A}} \, {\cal D} A \, {\cal D}
\bar{A} \exp i
\left( {\cal S}[A, \bar{A}] + \int d^4 x d^4 \theta (A \tilde{A} + \tilde{A}
\tilde{\tilde{A}}) + {\rm c.c.}
\right) \cr
& = \exp i {\cal S}[ - \tilde{\tilde{A}}, - \bar{\tilde{\tilde{A}}}].
}
}
This shows that $S^2 = 1$ on the action, up to the trivial change of variable
$A \rightarrow - A$.
Furthermore, a $T$-transformation followed by an $S$-transformation transform
an action functional ${\cal S}$ into ${\cal S}_1$ given by
\eqn\ST{
\exp i {\cal S}_1[A_1, \bar{A}_1] = \int {\cal D} A {\cal D} \bar{A} \exp i
\left( {\cal S} [A, \bar{A}] + \int d^4 x d^4 \theta ( {1 \over 2} A A + A A_1)
+ {\rm c.c.} \right).
}
If we now perform three consecutive $S T$-transformations, an action functional
${\cal S}$ is transformed into ${\cal S}_3$ given by
\eqn\STSTST{
\eqalign{
\exp i {\cal S}_3[A_3, \bar{A}_3] & = \int  {\cal D} A_2 \, {\cal D} \bar{A}_2
\, {\cal
D} A_1 \, {\cal D} \bar{A}_1 \, {\cal D} A \, {\cal D} \bar{A} \exp i \Bigl(
{\cal S}[A, \bar{A}] \cr
& \;\;\;\; \left. + \int d^4 x d^4 \theta ( {1 \over 2} A A
+ A A_1 + {1 \over 2} A_1 A_1 + A_1 A_2 + {1 \over 2} A_2 A_2 + A_2 A_3) + {\rm
c.c} \right) \cr
& = \exp i {\cal S} [A_3, \bar{A}_3]. \cr
}
}
Thus $(ST)^3 = 1$ on the action. The relations \relations\ being fulfilled,
the group generated by the $S$- and $T$-transformations is indeed isomorphic to
$SL(2, {\bf Z})$.

\newsec{The long-distance expansion}

The exact effective action ${\cal S}[A, \bar{A}]$ obtained by integrating out
all massive degrees of freedom from the microscopic theory is in general an
intractable
non-local expression. To be able to extract useful information about for
example a scattering process, we can expand this expression in powers of the
momentum scale of the external particles divided by the characteristic scale of
the theory and consider the leading terms. Since a space-time derivative in the
action corresponds to a power of space-time momentum, this procedure roughly
amounts to keeping terms with up to some maximal number of space-time
derivatives in the action.

We thus introduce an `order in derivatives' $n$ to the different objects in our
theory as follows: We define $n$ by specifying that the $N = 2$ vector field
$A$ has $n = 0$ and a super covariant derivative $D_\alpha{}^i$ has $n = {1
\over 2}$. ($n$ is always invariant under complex conjugation.) From the
anti-commutators of covariant derivatives \Dalgebra, it then follows that $n =
1$ for a space-time derivative $\partial_\mu$ as we wanted. The space-time
integration measure $d^4 x$ must  have $n = - 4$. From the explicit form of the
derivatives \susyderivatives, we see that the fermionic coordinates
$\theta^\alpha{}_i$ have $n = - {1 \over 2}$, and the rules of Grassmann
integration then give that $n = 2$ for the integration measure $d^4 \theta$.
Finally, we see that the delta functions $\delta^4(x)$ and $\delta^4(\theta)$
have $n = 4$ and $n = - 2$ respectively.

Decomposing $A$ into an $N = 1$ vector field $W_\alpha$ and an $N = 1$ chiral
field $\Phi$, we see that these have $n = {1 \over 2}$ and $n = 0$
respectively. The integration measure $d^2 \theta$ has $n = 1$. For the
component fields, we get $n = 0$ for the gauge potential $A_\mu$ and the scalar
field $\phi$, $n = {1 \over 2}$ for the spinor fields $\lambda_\alpha{}^i$, and
$n = 1$ for the auxiliary fields $E^A$. In particular, this means that the
canonical kinetic terms for these fields (i.e. the terms in the free Lagrangian
$i
\int d^4 \theta A A + {\rm c.c.}$) will all have $n = 2$. This, together with
the requirement that a space-time derivative has $n = 1$, could also be taken
as a (perhaps more intuitive) definition of $n$. (We caution the reader not to
confuse the order $n$ with the canonical dimension $d$. For example, $d = 1$
for the scalar field $\phi$ and the gauge potential $A_\mu$, $d = {3 \over 2}$
for the spinor fields $\lambda_\alpha{}^i$ and $d = 2$ for the auxiliary fields
$E^A$.)

In $N = 1$ supersymmetry, it is possible to write terms in the Lagrangian of
all integer orders $n$. The lowest order term in that case is the order $n = 1$
term
\eqn\potential{
\int d^2 \theta f(\Phi) + {\rm c.c.},
}
where the superpotential $f$ is an arbitrary holomorphic function. At order $n
= 2$ we have the terms
\eqn\Kahler{
\int d^2 \theta d^2 \bar{\theta} K(\Phi, \bar{\Phi})
}
and
\eqn\Maxwell{
\int d^2 \theta \tau(\Phi) W^\alpha W_\alpha + {\rm c.c.},
}
where the K\"ahler potential $K$ and the gauge coupling $\tau$ are arbitrary
real analytic and holomorphic functions respectively. The number of possible
terms then increases rapidly with the order $n$.

In $N = 2$ supersymmetry, only terms of even order $n$ are possible. At order
$n = 2$ we have
\eqn\prepotential{
\int d^4 \theta {\cal F}(A) + {\rm c.c.}
}
for an arbitrary holomorphic prepotential ${\cal F}$ \ref\Seibergtwo{N.
Seiberg,
{\it Phys. Lett.} {\bf 206B} (1988) 75.}. In terms of $N = 1$ superfields, this
term may be written as
\eqn\prepot{
\int d^2 \theta d^2 \bar{\theta} \bar{\Phi} {\cal F}^\prime(\Phi) + \int d^2
\theta {1 \over 2} {\cal F}^{\prime \prime}(\Phi) W^\alpha W_\alpha + {\rm
c.c.},
}
where a prime on a function denotes differentiation with respect to its
argument. At order $n = 4$ we have
\eqn\four{
\int d^4 \theta d^4 \bar{\theta} {\cal K}(A, \bar{A})
}
for an arbitrary real analytic function ${\cal K}$. In $N = 1$ superspace this
reads
\eqn\fyra{
\eqalign{
\int d^2 \theta d^2 \bar{\theta} & \Bigl( {\cal K}_{\phi \bar{\phi}} (\Phi,
\bar{\Phi}) \bigl( D^\alpha D_\alpha \Phi \bar{D}_{\dot{\alpha}}
\bar{D}^{\dot{\alpha}} \bar{\Phi} + 2 \bar{D}_{\dot{\alpha}} D^\alpha \Phi
D_\alpha \bar{D}^{\dot{\alpha}} \bar{\Phi} + 4 D^\alpha W_\alpha
\bar{D}_{\dot{\alpha}} \bar{W}^{\dot{\alpha}}  \cr
& \;\;\;\;\;\;\;\;\;\;\;\;\;\;\;\;\;\;\;\; - 4 D^{(\alpha} W^{\beta)}
D_{(\alpha} W_{\beta)} - 4  \bar{D}_{(\dot \alpha} \bar{W}_{\dot \beta)}
\bar{D}^{(\dot \alpha}  \bar{W}^{\dot \beta)} \cr
& \;\;\;\;\;\;\;\;\;\;\;\;\;\;\;\;\;\;\;\; - 2  D^\alpha D_\alpha (W^\beta
W_\beta) - 2 \bar{D}_{\dot{\alpha}} \bar{D}^{\dot{\alpha}} (
\bar{W}_{\dot{\beta}} \bar{W}^{\dot{\beta}} ) \bigr) \cr
& - 2 {\cal K}_{\phi \phi \bar{\phi}} (\Phi, \bar{\Phi}) W^\alpha W_\alpha
D^\beta D_\beta \Phi - 2 K_{\phi \bar{\phi} \bar{\phi}} (\Phi, \bar{\Phi})
\bar{W}_{\dot{\alpha}} \bar{W}^{\dot{\alpha}} \bar{D}_{\dot{\beta}}
\bar{D}^{\dot{\beta}} \bar{\Phi} \cr
& + K_{\phi \phi \bar{\phi} \bar{\phi}} (\Phi, \bar{\Phi}) \left( - 8 W^\alpha
D_\alpha \Phi \bar{W}_{\dot{\alpha}} \bar{D}^{\dot{\alpha}} \bar{\Phi} + 4
W^\alpha W_\alpha \bar{W}_{\dot{\alpha}} \bar{W}^{\dot{\alpha}} \right) \Bigr).
\cr
}
}
(The subscripts on the function ${\cal K}$ denote derivatives with respect to
its arguments.)
The expression \fyra\ is unique only up to total space-time derivatives or
terms that vanish because $D^\alpha W_\alpha = \bar{D}_{\dot{\alpha}}
\bar{W}^{\dot{\alpha}}$. The latter ambiguity makes the duality transformations
less straightforward in the $N = 1$ superspace formalism and is one of the
reasons that we prefer to work  in $N = 2$ superspace.

At order $n = 4$ a new phenomenon appears: If we expand out the Lagrangian
\four\ or \fyra\ in component fields, we will see that it contains a term
proportional to ${\cal K}_{\phi \bar{\phi}}(\phi, \bar{\phi}) \partial_\mu E^A
\partial^\mu E_A$. This means that the equations of motion for the auxiliary
fields $E^A$ will not be algebraic (unless ${\cal K}$ is a sum of a holomorphic
and an anti-holomorphic term, in which case the action vanishes). The auxiliary
fields are still non-propagating, though, because they may be solved for in the
equations of motion as an infinite series of terms in the other fields of all
orders $n$. (We recall that the auxiliary fields are `nominally' of order $n =
1$.) We remark that if we require only $N = 1$ supersymmetry it is possible to
find order $n = 4$ terms in the Lagrangian such that the auxiliary fields have
algebraic equations of motion, and this may be possible for $N = 2$
supersymmetry as well if we go to higher orders $n$. However, the solution for
the auxiliary fields to such equations of motion would still be of the same
form, i.e. an infinite series in the dynamical fields containing terms of all
orders $n$, so there is no reason to impose this extra requirement.

Finally, we will briefly consider the case of $N = 4$ supersymmetry. This is
less straightforward, since no off-shell formulation of $N = 4$ supersymmetry
is known. On-shell, an $N = 4$ vector multiplet can be decomposed as an $N = 1$
vector multiplet $W_\alpha$ and three $N = 1$ chiral multiplets $\Phi_a$, $a =
1, 2, 3$. The unique order $n = 2$ Lagrangian is
\eqn\fourtwo{
2 \tau \int d^2 \theta d^2 \bar{\theta} \bar{\Phi}^a \Phi_a + \tau \int d^2
\theta W^\alpha W_\alpha + {\rm c.c.},
}
where $\tau = {\theta \over 2 \pi} + i {4 \pi \over g^2}$. ($\theta$ and $g$
are the theta-angle and the gauge coupling constant respectively.) Putting the
auxiliary fields to zero by their equations of motion, the action is invariant
under the $SU(4)_R$ which is part of the $N = 4$ supersymmetry algebra. Only an
$SU(3) \times U(1)_R$ subgroup is manifest in \fourtwo, though, with $W_\alpha$
and $\Phi_a$ transforming in the ${\bf 1}_3$ and ${\bf 3}_2$  representations
respectively. To construct an order $n = 4$ term, we first consider the case
when $\Phi_2 = \Phi_3 = 0$ (assuming that this limit is smooth). The remaining
fields $W_\alpha$ and $\Phi = \Phi_1$ should then constitute an $N = 2$ vector
multiplet, and the Lagrangian must be of the form \fyra. Furthermore,
invariance under $U(1)_R$ requires ${\cal K}$ in \fyra\ to be a function of
$\Phi \bar{\Phi}$ only. We now reinstate $\Phi_2$ and $\Phi_3$ such that the
resulting Lagrangian is manifestly invariant also under $SU(3)$. Solving the
equations of motion for the auxiliary fields and substituting back in the
Lagrangian will only produce terms of order $n \geq 6$, so if we work at order
$n = 4$ we can put them to zero. Finally we must check that the remaining
Lagrangian is really invariant under $SU(4)$. This turns out to require that
${\cal K}(\Phi, \bar{\Phi}) = k \Phi \bar{\Phi}$ for some constant $k$.
Inserting this in \fyra\ and reinstating $\Phi_2$ and $\Phi_3$ gives the unique
term
\eqn\fourfour{
k \int d^2 \theta d^2 \bar{\theta} (2 D^\alpha D_\alpha \Phi_a
\bar{D}_{\dot{\alpha}} \bar{D}^{\dot{\alpha}} \bar{\Phi}^a - 8 D^\alpha
W_\alpha \bar{D}_{\dot{\alpha}} \bar{W}^{\dot{\alpha}})
}
up to a total space-time derivative and terms proportional to $D^\alpha
W_\alpha - \bar{D}_{\dot{\alpha}} \bar{W}^{\dot{\alpha}}$. Since the term
\fourfour\ is bilinear in the fields, we may in fact put the auxiliary fields
to zero to all orders $n$ by their equations of motion. It is conceivable that
terms in the Lagrangian of higher order $n$ depending on some arbitrary
functions, as in the case of $N = 1$ or $N = 2$ supersymmetry, may be
constructed. However, such functions would have to be real analytic rather than
holomorphic, since the only $SU(4)$ invariant combination of the scalar
component fields, namely $\phi_a \bar{\phi}^a$, is real.

\newsec{Transformation properties of  the effective action}

To determine the transformation properties of the different terms in the
effective action ${\cal S}$ under $SL(2, {\bf Z})$ duality, it is convenient to
decompose
it as
\eqn\action{
{\cal S}[A, \bar{A}] = \widehat{\cal S} [A, \bar{A}] + \int d^4 x d^4 \theta
{\cal F} (A) + {\rm c.c.},
}
where the first term contains all contributions to the action of order $n \geq
0$. (In this section, the order $n$ will always refer to terms in the action as
opposed to the Lagrangian, i.e. including the $n = -4$ contribution of the
space-time integration measure. The second term in \action\ thus has $n = -2$.)
Inserting \action\ in the formula \Saction\ for the dual action after an
$S$-transformation and decomposing this in analogy to \action, we get
\eqn\Sact{
\eqalign{
\exp i \tilde{\cal S} [\tilde{A}, \bar{\tilde{A}}] & = \exp i \left(
\widehat{\tilde{\cal S}} [\tilde{A}, \bar{\tilde{A}}] + \int d^4 x d^4 \theta
\tilde{\cal F} (\tilde{A}) +{\rm c.c.} \right) \cr
&  = \int {\cal D} A {\cal D} \bar{A} \exp i \left( \widehat{\cal S} [A,
\bar{A}] + \int d^4 x d^4 \theta ({\cal F} (A) + A \tilde{A}) + {\rm c.c.}
\right) \cr
& =  \int {\cal D} A {\cal D} \bar{A} \exp i \left( \widehat{\cal S} [ -i
{\delta \over \delta \tilde{A}}, -i {\delta \over \delta \bar{\tilde{A}}}] +
\int d^4 x d^4 \theta ({\cal F} (A) + A \tilde{A}) + {\rm c.c.} \right) \cr
& = \exp i \widehat{\cal S} [ -i {\delta \over \delta \tilde{A}}, -i {\delta
\over \delta \bar{\tilde{A}}}] \exp i \left( \int d^4 x d^4 \theta \tilde{\cal
F} (\tilde{A}) + {\rm c.c.} \right), \cr
}
}
where the dual prepotential $\tilde{\cal F}$ is given by
\eqn\Fdual{
\exp i \int d^4 x d^4 \theta \tilde{\cal F} (\tilde{A}) = \int {\cal D} A \exp
i \int d^4 x d^4 \theta ( {\cal F} (A) + A \tilde{A})
}
and the functional derivative is defined by
\eqn\functder{
{\delta A (x^\prime, \theta^\prime, \bar{\theta}^\prime) \over \delta A (x,
\theta, \bar{\theta})} = \bar{D}^{\prime 4} \left( \delta^4 (x - x^\prime)
\delta^4 (\theta - \theta^\prime) \delta^4 (\bar{\theta} - \bar{\theta}^\prime)
\right).
}
The four $\bar{D}^\prime$ operators on the right-hand side of \functder\ arise
because $A$ obeys the chirality constraint \chiral. We note that the operator
${\delta \over \delta A}$ is of order $n = 2$.

Although in general not a Gaussian, the path integral in \Fdual\ is not
difficult to evaluate. We require that the function $A_D$, defined by
\eqn\AD{
A_D = {\cal F}^\prime (A),
}
assumes every value exactly once. It then follows that, given $\tilde{A}$, the
integrand in \Fdual\ has a unique stationary point given by the equation ${\cal
F}^\prime (A) + \tilde{A} = 0$, i.e.
\eqn\ADAtilde{
\tilde{A} = - A_D.
}
We denote the corresponding value of $A$ as $\tilde{A}_D$ so that
\eqn\tildeAD{
{\cal F}^\prime (\tilde{A}_D) + \tilde{A} = 0.
}
We now claim that the path integral \Fdual\ is given by the value of the
integrand at this point, i.e. $\exp i \int d^4 x d^4 \theta \tilde{\cal F}
(\tilde{A}) = \exp i \int d^4 x d^4 \theta ( {\cal F}(\tilde{A}_D) +
\tilde{A}_D \tilde{A} )$. The dual prepotential is thus given by
\eqn\prepotdual{
\tilde{\cal F} (\tilde{A}) = {\cal F}(\tilde{A}_D) + \tilde{A}_D \tilde{A}.
}
The behavior of the integrand in \Fdual, apart from its value at the
stationary point, is of no consequence, since we can change it to any fiducial
function by a holomorphic change of variables, $A \rightarrow  y (A)$. It is
not difficult to check that the path integration measure ${\cal D} A$ is
invariant under such a transformation, the bosonic and fermionic contributions
to the Jacobian canceling because of supersymmetry. The path integral \Fdual\
thus equals the value of the integrand at the stationary point (times a
normalization constant which we can put to one).

Differentiating \prepotdual\ with respect to $\tilde{A}$ and using \tildeAD, we
get
\eqn\ADtilde{
\tilde{A}_D = \tilde{\cal F}^\prime (\tilde{A}),
}
which is the dual counterpart of \AD. Inserting this back in \tildeAD\ we get
${\cal F}^\prime (\tilde{\cal F}^\prime (\tilde{A})) = - \tilde{A}$. An
analogous formula appeared in the $N = 1$ superspace formalism in \SW. The
relationship, at the stationary point, between the variables before and after
dualization can be summarized as
\eqn\Smatrix{
\left( \matrix{A_D \cr A \cr} \right) \rightarrow \left( \matrix{\tilde{A}_D
\cr \tilde{A} \cr} \right) = \left( \matrix{0 & 1 \cr -1 & 0 \cr} \right)
\left( \matrix{A_D \cr A \cr} \right).
}

We now return our attention to the expression \Sact\ for the dual action.
Consider a term in $\widehat{\cal S}[A, \bar{A}]$ of some definite order $n$.
Replacing a power of $A$ by $- i {\delta \over \delta \tilde{A}}$ and letting
it act on an object of order $n^\prime$ would, by \functder, produce a result
of order $n + n^\prime + 2$. But $n^\prime \geq - 2$ with equality only for the
dual prepotential $\int d^4 x d^4 \theta \tilde{\cal F} (\tilde{A}) + {\rm
c.c.}$, so we see that a term of order $n$ in the action will only affect the
terms of equal or higher order in the dual action. The terms of equal order
arise when we let all functional derivatives act on the exponentiated dual
prepotential $\exp i \int d^4 x d^4 \theta \tilde{\cal F} (\tilde{A}) + {\rm
c.c.}$ so that $- i {\delta \over \delta \tilde{A}}$ can be replaced by
$\tilde{\cal F}^\prime (\tilde{A}) = \tilde{A}_D$. At the stationary point, we
can furthermore use \Smatrix\ to replace the latter expression by $A$.

In particular, consider the terms of order $n = 0$ in the dual action
(corresponding to the terms of order $n = 4$ in the Lagrangian) given by a real
analytic function $\tilde{\cal K}$ as we discussed in the previous section.
This function is completely determined by the terms of order $n \leq 0$ in the
original action, i.e. by the prepotential ${\cal F}$ and the real analytic
function ${\cal K}$. From the discussion in the previous paragraph, it follows
that
\eqn\Kdual{
\tilde{\cal K} (\tilde{A}, \bar{\tilde{A}}) = {\cal K} (A, \bar{A} ),
}
where by $A$ we understand its value at the stationary point of the integrand
in \Fdual\ as a function of $\tilde{A}$.

The $T$-transformation obviously acts as ${\cal F} (A) \rightarrow {\cal F} (A)
+ {1 \over 2} A A$, leaving $\widehat{\cal S} [A, \bar{A}]$ and in particular
${\cal K}(A, \bar{A})$ invariant. This can also be stated as
\eqn\Tmatrix{
 \left( \matrix{A_D \cr A \cr} \right) \rightarrow \left( \matrix{1 & 1 \cr 0 &
1 \cr} \right) \left( \matrix{A_D \cr A \cr} \right).
}

As in \SW, the interpretation of \Smatrix\ and \Tmatrix\ is that $\left(
\matrix{A_D \cr A \cr} \right)$ constitutes a section of an $SL(2, {\bf Z})$
bundle over the moduli space of vacua. Our result \Kdual\ concerning the terms
in the action of order $n = 0$ and the corresponding statement for the
$T$-transformation can then be summarized by saying that the real analytic
function ${\cal K}$ is a modular function with respect to $SL(2, {\bf Z})$. The
method of this section could in principle be applied to determine how the terms
in the action of any given order $n$ behave under duality transformations. (We
have already seen that their transforms are determined by terms of equal or
lower order in the original action.)

Much of the recent progress in supersymmetric gauge theories relies on the
holomorphicity of the superpotential in $N = 1$ theories or the prepotential in
$N = 2$ theories. When combined with a knowledge of singularity structure
and/or asymptotic behavior, this is often enough to completely determine these
functions. Unfortunately, we have seen that the higher order contributions to
the effective action are given by real analytic rather than holomorphic
objects, so they cannot be determined by this method. However, in the case of
$N = 2$ supersymmetry, the Coulomb branch of the moduli space of vacua can be
identified with a certain family of Riemann surfaces, and the prepotential is
then closely related to the periods of a certain one-form. It is conceivable
that also the higher order terms in the effective action, such as those
determined by the function ${\cal K}(A, \bar{A})$ that we have discussed, have
a geometric interpretation. We hope that the results of the present paper may
be helpful in clarifying this structure.

\bigskip I would like to thank G. Moore for discussions and the theory division
at CERN, where part of this work was done, for its hospitality. This research
was supported by DOE under grant DE-FG02-92ER40704.

\listrefs

\bye